\documentclass[aip,graphicx]{revtex4-1}
\usepackage{gensymb}
\usepackage{graphicx}
\usepackage{dcolumn}
\usepackage{bm}
\usepackage{wasysym}
\draft 

\begin{document}


\title{Temperature Dependence of Sensitivity of 2DEG-Based Hall-Effect Sensors} 



\author{H.S. Alpert}
\email[]{halpert@stanford.edu}

\affiliation{Department of Aeronautics and Astronautics, Stanford University, Stanford, CA 94305, USA}

\author{C.A. Chapin}
\affiliation{Department of Aeronautics and Astronautics, Stanford University, Stanford, CA 94305, USA}

\author{K.M. Dowling}
\affiliation{Department of Electrical Engineering, Stanford University, Stanford, CA 94305, USA}

\author{S.R. Benbrook}
\affiliation{Department of Electrical Engineering, Stanford University, Stanford, CA 94305, USA}

\author{H. K\"{o}ck}
\affiliation{Infineon Technologies Austria AG, Villach 9500, Austria}

\author{U. Ausserlechner}
\affiliation{Infineon Technologies Austria AG, Villach 9500, Austria}

\author{D.G. Senesky}
\affiliation{Department of Aeronautics and Astronautics, Stanford University, Stanford, CA 94305, USA}


\date{\today}

\begin{abstract}
The magnetic sensitivity of Hall-effect sensors made of InAlN/GaN and AlGaN/GaN heterostructures was measured between room temperature and 576\degree C. Both devices showed decreasing voltage-scaled magnetic sensitivity at high temperature, declining from 53 to 8.3~mV/V/T for the InAlN/GaN sample and from 89 to 8.5~mV/V/T for the AlGaN/GaN sample, corresponding to the decreasing electron mobility due to scattering effects at elevated temperatures. Alternatively, current-scaled sensitivities remained stable over the  temperature range, only varying by 13.1\% from the mean of 26.3~V/A/T and 10.5\% from the mean of 60.2~V/A/T for the InAlN/GaN and AlGaN/GaN samples respectively. This is due to the minimal temperature dependence of the electron sheet density on the 2-dimensional electron gas (2DEG). Both devices showed consistency in their voltage- and current-scaled sensitivity over multiple temperature cycles as well as nearly full recovery when returned to room temperature after thermal cycling. Additionally, an AlGaN/GaN sample held at 576\degree C for 12 hours also showed nearly full recovery at room temperature, further suggesting that GaN-based Hall-effect sensors are a good candidate for use in high temperature applications.
\end{abstract}


\maketitle 

\section{Introduction}
Hall-effect sensors are widely used in the automotive industry, in power electronics, and within inertial measurement units (IMUs) for navigation and position sensing. There is a growing need for Hall-effect sensors that can operate under extreme conditions, specifically in high temperature environments such as deep underground (e.g., well-logging) and in outer space. Applications within the space sector include current monitoring in hybrid rocket motors, power modules, and spacecraft motor control units \cite{krish14,koide12}. While many spacecraft experience moderately high temperatures due to solar heating and power dissipation, even higher temperatures must be endured during missions to planets in the inner Solar System like Venus, which has surface temperatures that regularly approach 500\degree C \cite{nasa14}. 

Electronic components, including Hall-effect sensors, are typically made of silicon due to its low cost, ease of manufacturing, and compatibility with integrated circuits. However, silicon-based components begin to breakdown at temperatures beyond 200\degree C \cite{lu06, koide12, hout97}, and thus external cooling is often required for electronics to operate in high temperature environments. However, implementing cooling processes requires additional power and contributes further bulk and complexity to the system, leading to increased size, weight, and overall costs of the system \cite{caltech14}. Thus, components that can operate at extreme temperatures without additional cooling are necessary for achieving higher efficiency, higher reliability, and lower cost.

Wide bandgap semiconductors such as gallium nitride (GaN) and aluminum nitride (AlN) have been shown to operate up to 1000\degree C in vacuum \cite{mai12} and thus are a prime candidate for electronics for space applications. GaN-based Hall-effect sensors have shown room temperature sensitivity and offset characteristics similar to those of silicon Hall-effect sensors \cite{alp19,dow19,vdm04,vdm05,aus16,ruth03,rand81,san15}, but also reliable operation up to 400\degree C \cite{koide12,boug09,abd12,lu06, wang07} for short periods of time. In this paper, we investigate how the sensitivities of InAlN/GaN and AlGaN/GaN Hall-effect sensors change with temperature (ranging from room temperature to 576\degree C), and further examine the change in sensitivity after exposing the GaN-based sensor to 576\degree C for a period of 12 hours.

\section{Device Microfabrication}

\begin{figure}
\includegraphics[width=\linewidth]{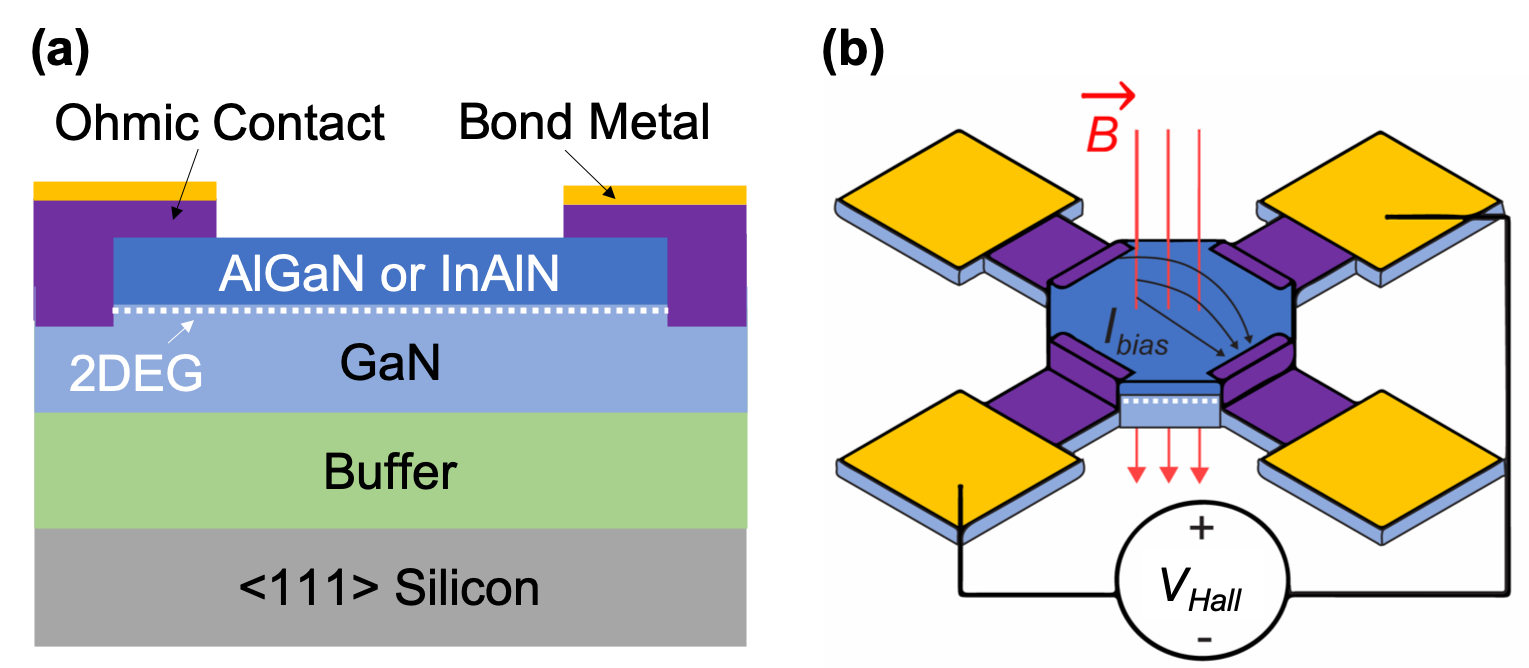}
\caption{\label{fig:hall}(a) Cross-sectional schematic and (b) operating principle of Hall-effect sensor.}
\end{figure}

The microfabrication process for the InAlN/GaN and AlGaN/GaN Hall-effect sensors was similar to that previously reported by us elsewhere \cite{alp19}, and a brief description is provided here for clarity. The InAlN/GaN-on-silicon wafer was purchased from NTT Advanced Technology Corporation, and the AlGaN/GaN was grown by metal-organic chemical vapor deposition (MOCVD) on a silicon wafer in the Stanford Nanofabrication Facility. For both wafers, an etch was performed on the III-nitride layer to isolate the mesa, a metal stack of Ti (20 nm)/Al (200 nm)/Mo (40 nm)/Au (80 nm) was deposited and annealed at 850\degree C for 35 seconds to form Ohmic contacts, and bond metal (Ti/Al) was deposited on top to allow for wire bonding. Unlike the process described in Ref.~\onlinecite{alp19}, no passivation layer was deposited on these samples. A cross-sectional schematic of the Hall plate is shown in Fig. \ref{fig:hall}a.

The InAlN/GaN and AlGaN/GaN sensors tested in this study were octagonal Hall plates. Those tested for the initial sensitivity sweep were regular octagons, where the sides with and without contacts are of equal lengths. The InAlN/GaN sample tested had dimensions $d=100$~$\mu$m and $a=41.4$~$\mu$m, where $d$ is the distance between transverse contacts and $a$ is the length of the sides with contacts, while the AlGaN/GaN sample tested had dimensions $d=200$~$\mu$m and $a=82.8$~$\mu$m. Meanwhile, the Hall-effect plate used for the 12-hour high temperature test had “longer” contacts; in this device, the contact length was 2.33 times as long as the sides without contacts and the dimensions were $d=200$~$\mu$m and $a=124.4$~$\mu$m (described further in Ref.~\onlinecite{alp19}).

\section{Experimental}

\begin{figure*}
\includegraphics[width=\linewidth]{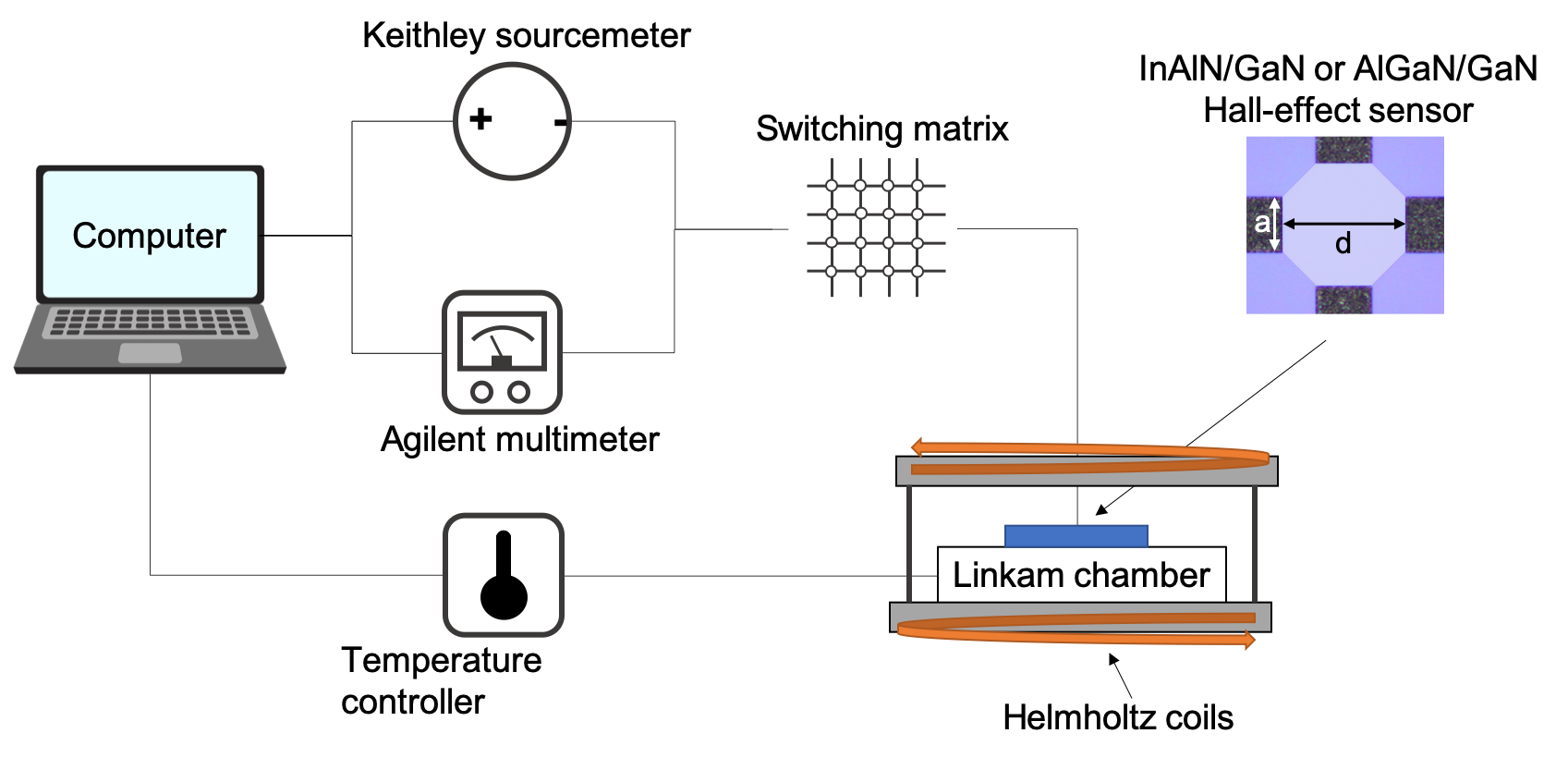}
\caption{\label{fig:testsetup}Diagram of experimental setup. During testing, the Hall-effect sensor was epoxied to an aluminum sheet placed atop the heating stage, and the device was wirebonded directly to pins connecting to electrical feedthroughs.}
\end{figure*}

\subsection{Principle of Operation}
The principle of operation of Hall-effect sensors is shown in Fig. \ref{fig:hall}b. Determining the sensitivity of a Hall-effect sensor requires measuring the Hall voltage ($V_{H}$), which is defined as
\begin{equation}
V_{H} = \frac{IBr_{n}G_{H}}{qn_{s}},
\end{equation}
where $I$ is the applied current, $B$ is the external magnetic field, $q$ is the electronic charge, and $n_{s}$ is the sheet electron density of the two-dimensional electron gas (2DEG). This equation also includes two proportionality constants: $r_{n}$ is the material-based scattering factor, which has been shown to be $\sim$1.1 in GaN \cite{rid00}, and $G_{H}$ is the geometry-dependent shape factor that accounts for the short-circuiting effects of having finite contacts (e.g., reduction in Hall voltage, change in linearity) \cite{aus17,rand81,popo04}. 

The sensitivity of a Hall-effect device with respect to supply current ($S_i$) is inversely proportional to $n_{s}$;
\begin{equation}
S_{i} = \frac{V_{H}}{IB} = \frac{r_{n}}{qn_{s}}G_{H}.
\end{equation}
Reducing the sheet electron density of the 2DEG causes an increase in the sheet resistance; a constant supply voltage will correspond to a lower supply current, therefore increasing the current-scaled sensitivity of the device.

In addition, the sensitivity with respect to supply voltage ($S_v$) is proportional to electron mobility ($\mu_H$); 
\begin{equation}
S_{v} = \frac{V_{H}}{V_sB} = \frac{r_{n}G_{H}}{Rqn_{s}} = \mu_Hr_{n}\frac{G_{H}}{{(\frac{L}{W})}_{eff}},
\end{equation}
where $V_s$ is supply voltage, $R$ is the device resistance, and $(L/W)_{eff}$ is the effective number of squares, defined as the ratio of the internal resistance to the sheet resistance \cite{aus16_2}.

\subsection{Test Procedures}

To conduct the sensitivity tests, the devices were diced into square dies with side lengths of $\sim$2~mm and epoxied with Durabond 952 Epoxy to a 1~in.~$\times$~1~in. aluminum sheet with a thickness of 1~mm, which was subsequently placed on a heating stage manufactured by Linkam Scientific Instruments. The contacts were wirebonded directly to the electrical connections of the chamber, which were then connected to a sourcemeter (Kiethley 2400) to generate a voltage between two contacts, and a multimeter (Agilent 34410A) to measure the Hall voltage generated across the two transverse contacts. A switching matrix (U2715A) was used to implement current spinning, by alternating the source and sense contacts between eight configurations (described in Ref.~\onlinecite{alp19}). The heating stage was placed between two copper coils wound around a 3D-printed scaffold, and current was applied through the coils to generate a magnetic field of 2~mT. A diagram of the test setup is shown in Figure \ref{fig:testsetup}.

For the initial sensitivity sweep, the device was supplied with three different bias voltages (0.3~V, 0.5~V, and 1~V), and ten measurements were taken under each bias condition. Measurements were first taken at room temperature and then subsequently at higher temperatures in steps of 25-50\degree C, until reaching 576\degree C. The temperature was then ramped back down to room temperature. The InAlN/GaN sample underwent two temperature cycles, while the process was repeated a third time for the AlGaN/GaN sample. 

For the 12-hour high temperature test, the same measurements (10 measurements at each of three supply voltages) were taken at room temperature, and then the device was held at 576\degree C for 12 hours and subsequently returned to room temperature, where the measurements were taken once again.

Although the heating stage itself is capable of reaching 600\degree C, there was a substantial difference between the temperature of the stage and that of the device under test at high temperatures. To characterize the true temperature of the device throughout the experiment, a resistance temperature detector (RTD) was integrated with a Hall plate during a temperature sweep. The temperature difference between the chuck and the device was additionally confirmed during a temperature sweep up to 200\degree C in which a thermocouple was epoxied to the aluminum sheet in the same manner as the devices under test. The temperature readouts between the thermocouple and the RTD matched to within 1.3\% and thus are the temperatures reported here.

\begin{figure*}
\includegraphics[width=\linewidth]{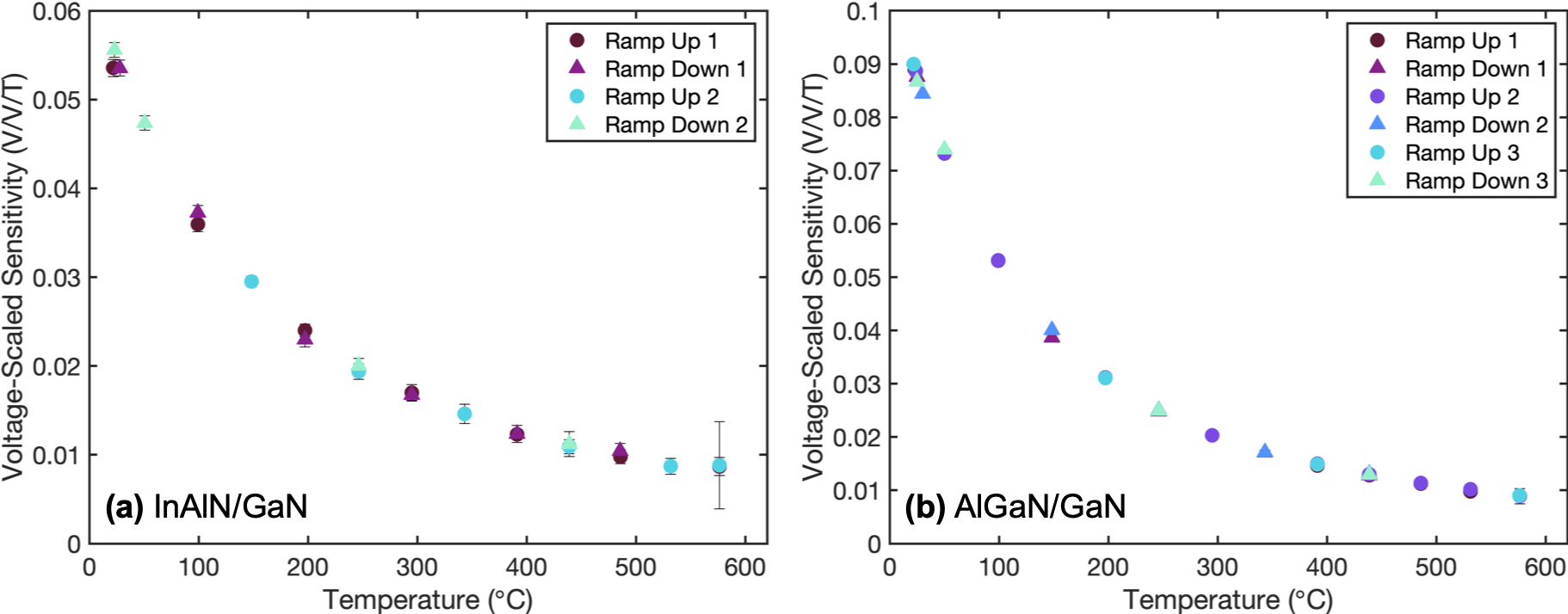}
\caption{\label{fig:V_sen}Voltage-scaled sensitivity of (a) InAlN/GaN and (b) AlGaN/GaN samples between room temperature and 576$^\circ$C.}
\end{figure*}

\section{Results and Discussion}

\begin{figure}
\includegraphics[width=4in]{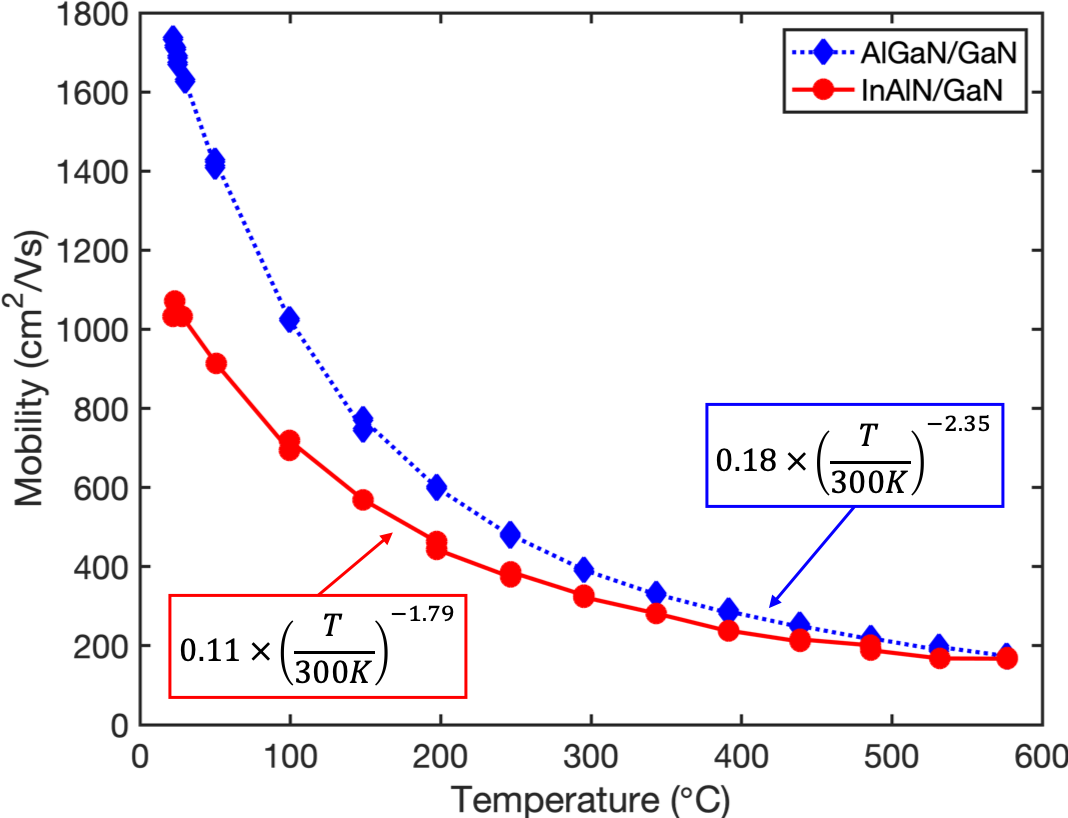}
\caption{\label{fig:mobility}Mobility of AlGaN/GaN and InAlN/GaN samples between room temperature and 576$^\circ$C.}
\end{figure}

Fig. \ref{fig:V_sen} plots the voltage-scaled sensitivity from room temperature to 576\degree C for the InAlN/GaN and AlGaN/GaN devices with a bias voltage of 0.3~V. The voltage-scaled sensitivities dropped from 53 to 8.3~mV/V/T for the InAlN/GaN sample and from 89 to 8.5~mV/V/T for the AlGaN/GaN sample over the temperature range. The mobility of the 2DEG was calculated from Equation 3, using the voltage-scaled sensitivity, geometry factor, and the scattering factor of the device. Fig. \ref{fig:mobility} plots the temperature dependence of the mobility for the InAlN/GaN and AlGaN/GaN Hall-effect sensors. The devices have a room temperature mobility of 1052~cm\textsuperscript{2}/V$\cdot$s and 1704~cm\textsuperscript{2}/V$\cdot$s for the InAlN/GaN and AlGaN/GaN respectively, and a mobility of 169~cm\textsuperscript{2}/V$\cdot$s and 172~cm\textsuperscript{2}/V$\cdot$s at 576\degree C. The mobility decrease with increasing temperature follows a power law of $0.11\times(T/300K)^{-1.79}$ for the InAlN/GaN sample and $0.18\times(T/300K)^{-2.35}$ for the AlGaN/GaN sample, which agrees with many results published in literature \cite{amin16}. The decline in voltage-scaled sensitivity at high temperature is largely due to the corresponding decrease in electron mobility, caused by increased scattering at high temperatures \cite{mnat03}.

\begin{figure*}
\includegraphics[width=\linewidth]{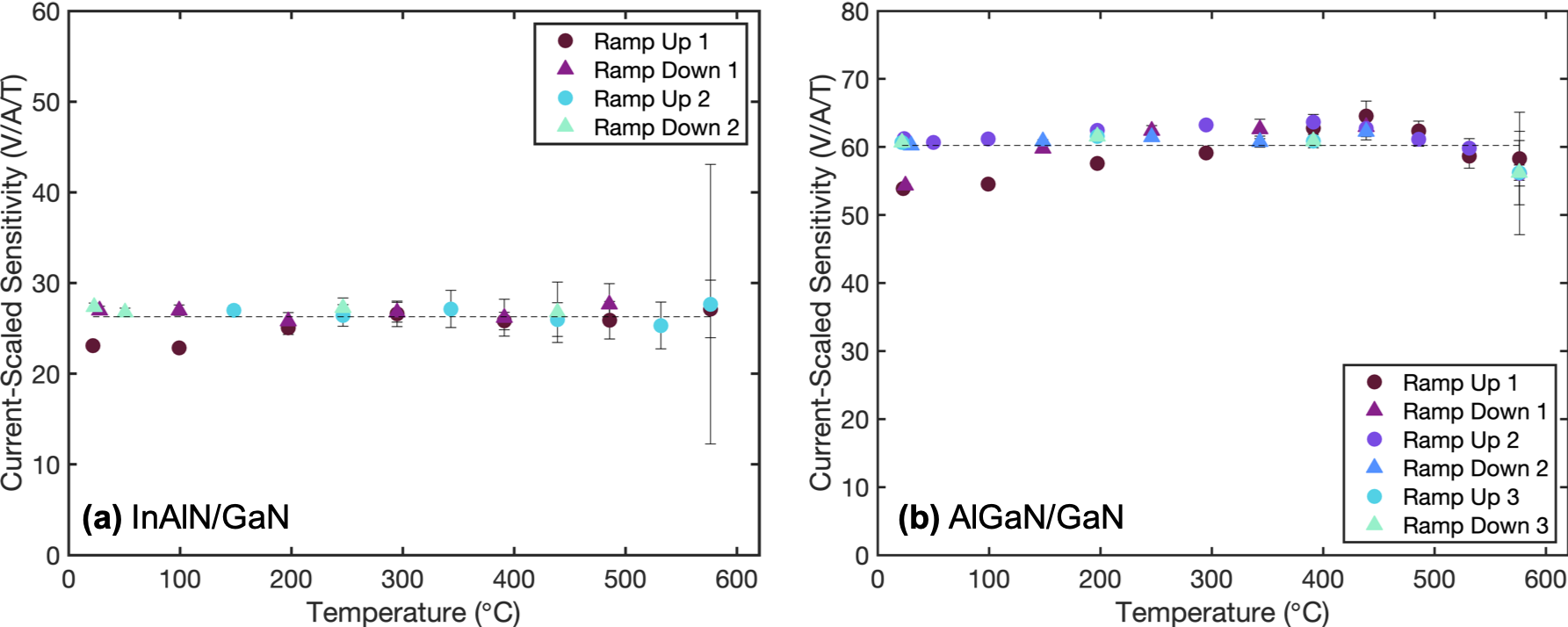}
\caption{\label{fig:I_sen}Current-scaled sensitivity of (a) InAlN/GaN and (b) AlGaN/GaN samples between room temperature and 576$^\circ$C.}
\end{figure*}

\begin{figure}
\includegraphics[width=4in]{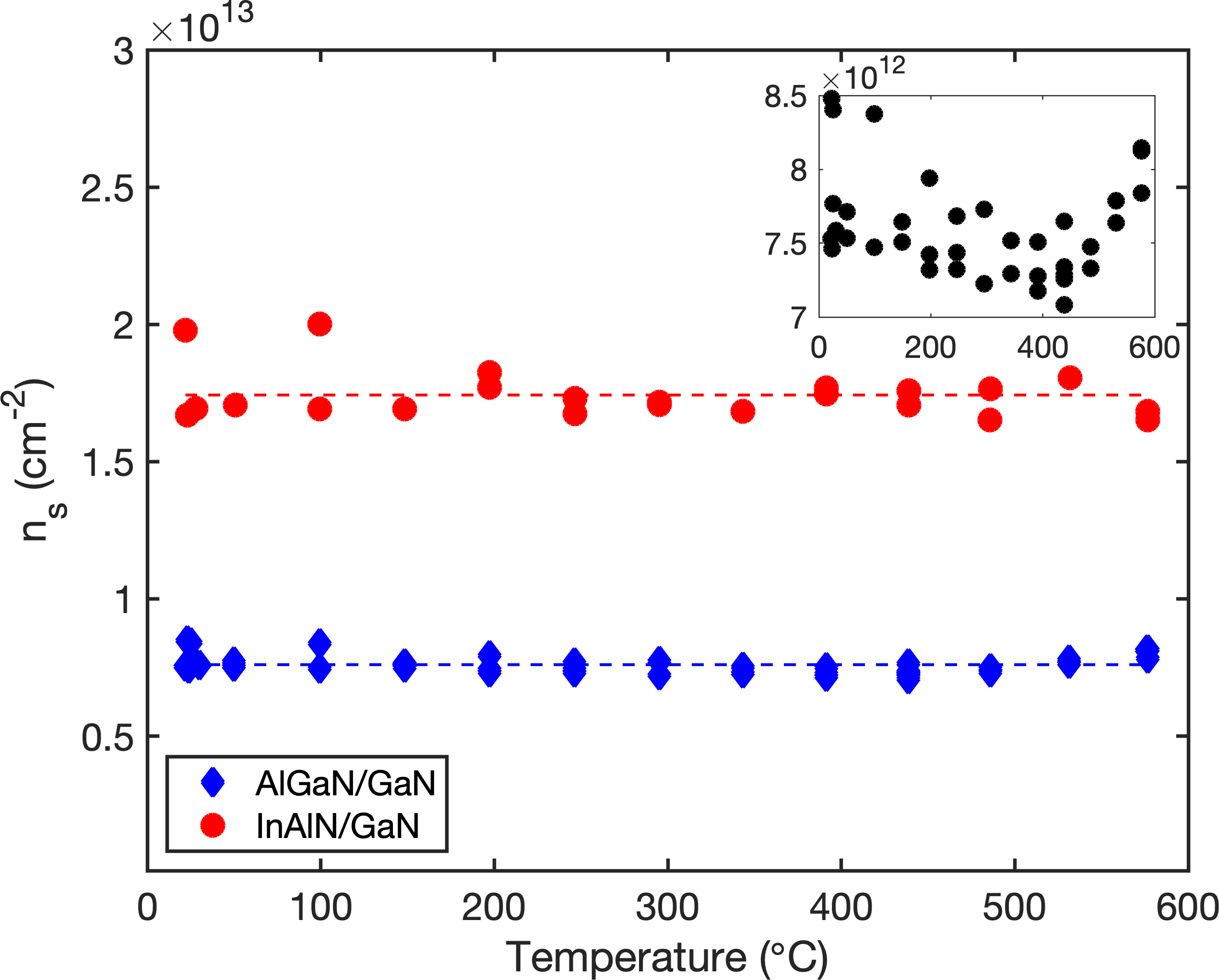}
\caption{\label{fig:2DEG} 2DEG sheet density of InAlN/GaN and AlGaN/GaN samples as a function of temperature from room temperature to 576$^\circ$C. The inset shows the AlGaN/GaN 2DEG sheet density over temperature on tighter axes.}
\end{figure}

The current-scaled sensitivity of the two devices stayed relatively constant with temperature (Fig. \ref{fig:I_sen}), indicating a stable 2DEG sheet density (Fig. \ref{fig:2DEG}). The current-scaled sensitivity varied by 13.1\% from the mean of 26.3~V/A/T for the InAlN/GaN sample and 10.5\% from the mean of 60.2~V/A/T for the AlGaN/GaN sample over the whole temperature range. Upon closer examination, during the first temperature ramp, the AlGaN/GaN showed a decrease in 2DEG density until about 350-400\degree C and then subsequently an increase (inset of Fig. \ref{fig:2DEG}). This same behavior was described in Ref.~\onlinecite{wang07} and attributed to conduction band lowering. However, in the subsequent temperature ramps this profile flattened out dramatically, suggesting that conduction band lowering may not actually be the sole cause of this behavior.

Another change that appeared to take place between the first temperature cycle and the ensuing cycles was the current-scaled sensitivity at room temperature. For both material platforms, the current-scaled sensitivity was at its minimum the first time it was measured, and it then increased in following cycles. While we at first attributed this to a permanent change in the material (e.g., thermally-induced strain), conducting further testing revealed this same behavior (having the lowest sensitivity at the start of the first thermal cycle) many days later. One possible explanation is that moisture that accumulated on the device was burned off during the first temperature cycle, temporarily changing the device resistance \cite{aus04}. 

Sensitivity measurements taken before and after storing an AlGaN/GaN device at 576\degree C for 12 hours showed the ability of the sensor to survive extreme temperatures for an extended period of time. The voltage-scaled sensitivity changed from 92~mV/V/T before being subjected to high temperature to 86~mV/V/T afterward, while the current-scaled sensitivity changed from 38.9~V/A/T to 39.7~V/A/T. Thus, the voltage- and current-scaled sensitivities shifted by -6.5\% and 2.6\% respectively, suggesting nearly full recovery. The variation between the ten measurements taken before the thermal storage was less than 1\% for both sensitivity metrics, as was also the case for the ten measurements taken after thermal storage.

\section{Conclusions}
We tested the sensitivity of InAlN/GaN and AlGaN/GaN Hall-effect plates across a range of temperatures between room temperature and 576\degree C. Both material platforms showed decreasing voltage-scaled sensitivity with increasing temperature, proportional to the decrease in mobility due to scattering effects. The devices showed relatively stable current-scaled sensitivities across the temperature range, suggesting a stable 2DEG sheet density. Additionally, the ability of the AlGaN/GaN sensor to survive 12 hours at 576\degree C and recover to nearly its original sensitivity values after returning to room temperature suggests the durability of this material for extreme environment applications. Further work will examine the behavior of these devices as they are subjected to high temperatures for longer periods of time, to more closely simulate the conditions of space missions and quantify the reliability of the sensors through lifetime testing. Finally, we plan to test these sensors in other extreme aspects of the space environment (e.g., gamma radiation) to further investigate their potential for use in space applications.

\begin{acknowledgments}
This work was supported in part by the Stanford SystemX Alliance, the National Defense Science and Engineering Graduate Fellowship, and the National Science Foundation Engineering Research Center for Power Optimization of Electro-Thermal Systems (POETS) with cooperative agreements EEC-1449548. Fabrication work was performed in part at the Stanford Nanofabrication Facility (SNF) and the Stsanford Nano Shared Facilities (SNSF), supported by the National Science Foundation under award ECCS-1542152.
\end{acknowledgments}

%
%

%


\bibliography{HighTempPaper}

@PREAMBLE{
 "\providecommand{\noopsort}[1]{}" 
 # "\providecommand{\singleletter}[1]{#1}
}


@ARTICLE{krish14,
   author       = "A. Krishna S. and L. Abraham", 
   title        = "Analysis of Different Hall Effect Current Sensors for Space Applications",
   journal      = "International J. of Innovative Sci., Eng., Tech.", 
   volume       = "1", 
   number       = "5",
   pages        = "380--386", 
   year         = "2014", 
}

@inproceedings{koide12,
  title="High Temperature Hall sensors using {A}l{G}a{N}/{G}a{N} {HEMT} Structures",
  author="S. Koide and H. Takahashi and A. Abderrahmane and I. Shibasaki, A. and A. Sandhu",
  booktitle="J. of Physics: Conf. Ser.",
  volume = "352",
  pages={879--888},
  year={2011},
  address="",
}


@TECHREPORT{nasa14,
  AUTHOR =        {},
  TITLE =         {Venus Technology Plan},
  INSTITUTION =   {NASA},
  MONTH =         {May},
  YEAR  =         {2014},
  PAGES =         {28},

}

@TECHREPORT{caltech14,
  AUTHOR =        "T. S. Balint and J. A. Cutts and E. A. Kolawa and C. E. Peterson",
  TITLE =         "Extreme Environment Technologies for Space and Terrestrial Applications",
  INSTITUTION =   "JPL, California Institute of Technology",
  MONTH =         {},
  YEAR  =         {2014},
  PAGES =         {12},

}

@ARTICLE{lu06,
   author       = "H. Lu and P. Sandvik and A. Vertiatchikh and J. Tucker and A. Elasser", 
   title        = "High temperature Hall effect sensors based on {A}l{G}a{N}/{G}a{N} heterojunctions",
   journal      = "J. App. Phys.", 
   volume       = "99", 
   pages        = "114510-1--114510-4", 
   year         = "2006", 
}

@ARTICLE{hout97,
   author       = "S. R. Hout and S. Middelhoek", 
   title        = "A 400\degree {C} silicon Hall sensor",
   journal      = "Sensors and Actuators A", 
   volume       = "60", 
   pages        = "14--22", 
   year         = "1997", 
}


@ARTICLE{mai12,
   author       = "D. Maier and M. Alomari and N. Grandjean and J.-F. Carlin and M.-A. Diforte-Poisson and C. Dua and S. Delage and E. Kohn", 
   title        = "In{A}l{N}/{G}a{N} {HEMT}s for Operation in the 1000\degree {C} Regime: A First Experiment",
   journal      = "IEEE Electron Dev. Lett.", 
   volume       = "33", 
   number        = "7",
   pages        = "985--987", 
   year         = "2012", 
}

@ARTICLE{alp19,
   author       = "H. S. Alpert and K. M. Dowling and C. A. Chapin and A. S. Yalamarthy and S. R. Benbrook and H. K{\"{o}}ck and U. Ausserlechner and D. G. Senesky", 
   title        = "Effect of Geometry on Sensitivity and Offset of {A}l{G}a{N}/{G}a{N} and {I}n{A}l{N}/{G}a{N} Hall-Effect Sensors",
   journal      = "IEEE Sensors J.", 
   volume       = "19", 
   number       = "10",
   pages        = "3640--3646", 
   year         = "2019", 
}

@ARTICLE{dow19,
   author       = "K. M. Dowling and H. S. Alpert and A. S. Yalamarthy and P. F. Satterthwaite and S. Kumar and H. K{\"{o}}ck and U. Ausserlechner and D. G. Senesky", 
   title        = "Micro-Tesla Offset in Thermally Stable {A}l{G}a{N}/{G}a{N} {2DEG} Hall Plates Using Current Spinning",
   journal      = "IEEE Sensors Lett.", 
   volume       = "3", 
   number       = "3",
   pages        = "", 
   year         = "2019", 
}

@inproceedings{vdm04,
  title="{CMOS} quad spinning-current Hall sensor system for compass application",
  author="J. C. van der Meer and F. R. Riedijik and P. C. de Jong and E. A. van Kampen and M. J. Meekel and J. H. Huijsing",
  booktitle="Proc. Int. Conf. IEEE Sensors",
  volume = "",
  pages="1434--1437",
  year={2004},
  address="",
}

@inproceedings{vdm05,
  title="A fully integrated {CMOS} Hall sensor with a 3.65 $\mu${T} 3 $\sigma$ offset for compass applications",
  author="J. C. van der Meer and F. R. Riedijik and E. van Kampen and K. A. A. Makinwa and J. H. Huijsing",
  booktitle="Proc. Int. Conf. IEEE Sensors",
  volume = "",
  pages="246--247",
  year={2005},
  address="San Francisco, CA, USA",
}

@ARTICLE{aus16,
   author       = "U. Ausserlechner", 
   title        = "Hall Effect Devices with Three Terminals: Their Magnetic Sensitivity and Offset Cancellation Scheme",
   journal      = "J. of Sensors", 
   volume       = "2016", 
   number       = "",
   pages        = "", 
   year         = "2016", 
}

@inproceedings{san15,
  title="Isotropic 3{D} silicon Hall Sensor",
  author="C. Sander and C. Leube and T. Aftab and P. Ruther and O. Paul",
  booktitle="Proc. 28th Int. Conf. IEEE Micro. Electro. Mech. Syst.",
  volume = "",
  pages="893--896",
  year={2015},
  address="Estoril, Portugal",
}

@ARTICLE{rand81,
   author       = "G. S. Randhawa", 
   title        = "Monolithic integrated Hall devices in silicon circuits",
   journal      = "Microelectronics J.", 
   volume       = "12", 
   number       = "6",
   pages        = "24--29", 
   year         = "1981", 
}

@ARTICLE{ruth03,
   author       = "P. Ruther and U. Schiller and R. Janke and O. Paul", 
   title        = "Thermomagnetic residual offset in integrated Hall plates",
   journal      = "IEEE Sensors J.", 
   volume       = "3", 
   number       = "6",
   pages        = "693--699", 
   year         = "2003", 
}

@ARTICLE{abd12,
   author       = "A. Abderrahmane and S. Koide and S.-I. Sato and T. Ohshima and A. Sandhu and H. Okada", 
   title        = "Robust Hall Effect Magnetic Field Sensors for Operation at High Temperatures and in Harsh Radiation Environments",
   journal      = "IEEE Trans. on Magnetics", 
   volume       = "48", 
   number       = "11",
   pages        = "4421--4423", 
   year         = "2012", 
}

@ARTICLE{boug09,
   author       = "L. Bouguen and L. Konczewicz and S. Contreras and B. Jouault and J. Camassel and Y. Cordier", 
   title        = "High temperature behaviour of {A}l{G}a{N}/{G}a{N} Hall-{FET} sensors",
   journal      = "Materials Sci. and Eng. B", 
   volume       = "165", 
   number       = "",
   pages        = "1--4", 
   year         = "2009", 
}

@ARTICLE{wang07,
   author       = "M. J. Wang and B. Shen and F. J. Xu and Y. Wang and J. Xu and S. Huang and Z. J. Yang and K. Xu and G. Y. Zhang", 
   title        = "High temperature dependence of the density of two-dimensional electron gas in {A}l$_{0.18}${G}a$_{0.82}${N}/{G}a{N} heterostructures",
   journal      = "Appl. Phys. A", 
   volume       = "", 
   number       = "",
   pages        = "715--718", 
   year         = "2007", 
}

@ARTICLE{jan11,
   author       = "J. Jankowski and S. El-Ahmar and M. Oszwalkdowski", 
   title        = "Hall sensors for extreme temperatures",
   journal      = "Sensors (Basel)", 
   volume       = "11", 
   number       = "1",
   pages        = "876--885", 
   year         = "2011", 
}

@ARTICLE{aus17,
   author       = "U. Ausserlechner", 
   title        = "The signal-to-noise ratio and a hidden symmetry of Hall plates",
   journal      = "Solid-State Electron.", 
   volume       = "135", 
   number       = "",
   pages        = "14--23", 
   year         = "2017", 
}

@book{popo04,
  author    = "R. Popovi\'{c}", 
  title     = "Hall Effect Devices",
  publisher = {IOP},
  year      = 2004,
  edition   = 2,
  address   = "UK",
}

@ARTICLE{rid00,
   author       = "B. K. Ridley and B. E. Foutz and L. F. Eastman", 
   title        = "Mobility of electrons in bulk {G}a{N} and {A}l$_x${G}a$_{1-x}${N}/{G}a{N} heterostructures",
   journal      = "Phys. Rev. B, Condens. Matter", 
   volume       = "611", 
   number       = "24",
   pages        = "16862--16869", 
   year         = "2000", 
}

@ARTICLE{aus16_2,
   author       = "U. Ausserlechner", 
   title        = "Closed form expressions for sheet resistance and mobility from Van-der-Pauw measurement on 90\degree symmetric devices with four arbitrary contacts",
   journal      = "Solid-State Electron.", 
   volume       = "116", 
   number       = "",
   pages        = "46--54", 
   year         = "2016", 
}

@ARTICLE{mnat03,
   author       = "T. T. Mnatsakanov and M. E. Levinshtein and L. I. Pomortseva and S. N. Yurkov and G. S. Simin and M. A. Khan", 
   title        = "Carrier mobility model for {G}a{N}",
   journal      = "Solid-State Electron.", 
   volume       = "47", 
   number       = "",
   pages        = "111--115", 
   year         = "2003", 
}

@ARTICLE{amin16,
   author       = "A. Aminbeidokhti and S. Dimitrijev and J. Han and X. Chen and X. Xu", 
   title        = "The Power Law of Phonon-Limited Electron Mobility in the 2-{D} Electron Gas of {A}l{G}a{N}/{G}a{N} Heterostructure",
   journal      = "IEEE Trans. Electron. Dev.", 
   volume       = "63", 
   number       = "5",
   pages        = "2214--2218", 
   year         = "2016", 
}

@inproceedings{aus04,
  title="Drift of magnetic sensitivity of smart {H}all sensors due to moisture absorbed by the {IC}-package [automotive applications]",
  author="U. Ausserlechner and M. Motz and M. Holliber",
  booktitle="SENSORS",
  volume = "",
  pages="455--458",
  year={2004},
  address="Vienna, Austria",
}

\providecommand{\noopsort}[1]{}\providecommand{\singleletter}[1]{#1}%
\begin{thebibliography}{25}%
\makeatletter
\providecommand \@ifxundefined [1]{%
 \@ifx{#1\undefined}
}%
\providecommand \@ifnum [1]{%
 \ifnum #1\expandafter \@firstoftwo
 \else \expandafter \@secondoftwo
 \fi
}%
\providecommand \@ifx [1]{%
 \ifx #1\expandafter \@firstoftwo
 \else \expandafter \@secondoftwo
 \fi
}%
\providecommand \natexlab [1]{#1}%
\providecommand \enquote  [1]{``#1''}%
\providecommand \bibnamefont  [1]{#1}%
\providecommand \bibfnamefont [1]{#1}%
\providecommand \citenamefont [1]{#1}%
\providecommand \href@noop [0]{\@secondoftwo}%
\providecommand \href [0]{\begingroup \@sanitize@url \@href}%
\providecommand \@href[1]{\@@startlink{#1}\@@href}%
\providecommand \@@href[1]{\endgroup#1\@@endlink}%
\providecommand \@sanitize@url [0]{\catcode `\\12\catcode `\$12\catcode
  `\&12\catcode `\#12\catcode `\^12\catcode `\_12\catcode `\%12\relax}%
\providecommand \@@startlink[1]{}%
\providecommand \@@endlink[0]{}%
\providecommand \url  [0]{\begingroup\@sanitize@url \@url }%
\providecommand \@url [1]{\endgroup\@href {#1}{\urlprefix }}%
\providecommand \urlprefix  [0]{URL }%
\providecommand \Eprint [0]{\href }%
\providecommand \doibase [0]{http://dx.doi.org/}%
\providecommand \selectlanguage [0]{\@gobble}%
\providecommand \bibinfo  [0]{\@secondoftwo}%
\providecommand \bibfield  [0]{\@secondoftwo}%
\providecommand \translation [1]{[#1]}%
\providecommand \BibitemOpen [0]{}%
\providecommand \bibitemStop [0]{}%
\providecommand \bibitemNoStop [0]{.\EOS\space}%
\providecommand \EOS [0]{\spacefactor3000\relax}%
\providecommand \BibitemShut  [1]{\csname bibitem#1\endcsname}%
\let\auto@bib@innerbib\@empty
\bibitem [{\citenamefont {S.}\ and\ \citenamefont {Abraham}(2014)}]{krish14}%
  \BibitemOpen
  \bibfield  {author} {\bibinfo {author} {\bibfnamefont {A.~K.}\ \bibnamefont
  {S.}}\ and\ \bibinfo {author} {\bibfnamefont {L.}~\bibnamefont {Abraham}},\
  }\bibfield  {title} {\enquote {\bibinfo {title} {Analysis of different hall
  effect current sensors for space applications},}\ }\href@noop {} {\bibfield
  {journal} {\bibinfo  {journal} {International J. of Innovative Sci., Eng.,
  Tech.}\ }\textbf {\bibinfo {volume} {1}},\ \bibinfo {pages} {380--386}
  (\bibinfo {year} {2014})}\BibitemShut {NoStop}%
\bibitem [{\citenamefont {Koide}\ \emph {et~al.}(2011)\citenamefont {Koide},
  \citenamefont {Takahashi}, \citenamefont {Abderrahmane}, \citenamefont
  {I.~Shibasaki},\ and\ \citenamefont {Sandhu}}]{koide12}%
  \BibitemOpen
  \bibfield  {author} {\bibinfo {author} {\bibfnamefont {S.}~\bibnamefont
  {Koide}}, \bibinfo {author} {\bibfnamefont {H.}~\bibnamefont {Takahashi}},
  \bibinfo {author} {\bibfnamefont {A.}~\bibnamefont {Abderrahmane}}, \bibinfo
  {author} {\bibfnamefont {A.}~\bibnamefont {I.~Shibasaki}}, \ and\ \bibinfo
  {author} {\bibfnamefont {A.}~\bibnamefont {Sandhu}},\ }\bibfield  {title}
  {\enquote {\bibinfo {title} {High temperature hall sensors using
  {A}l{G}a{N}/{G}a{N} {HEMT} structures},}\ }in\ \href@noop {} {\emph {\bibinfo
  {booktitle} {J. of Physics: Conf. Ser.}}},\ Vol.\ \bibinfo {volume} {352}\
  (\bibinfo {year} {2011})\ pp.\ \bibinfo {pages} {879--888}\BibitemShut
  {NoStop}%
\bibitem [{nas(2014)}]{nasa14}%
  \BibitemOpen
  \href@noop {} {\enquote {\bibinfo {title} {Venus technology plan},}\
  }\bibinfo {type} {Tech. Rep.}\ (\bibinfo  {institution} {NASA},\ \bibinfo
  {year} {2014})\BibitemShut {NoStop}%
\bibitem [{\citenamefont {Lu}\ \emph {et~al.}(2006)\citenamefont {Lu},
  \citenamefont {Sandvik}, \citenamefont {Vertiatchikh}, \citenamefont
  {Tucker},\ and\ \citenamefont {Elasser}}]{lu06}%
  \BibitemOpen
  \bibfield  {author} {\bibinfo {author} {\bibfnamefont {H.}~\bibnamefont
  {Lu}}, \bibinfo {author} {\bibfnamefont {P.}~\bibnamefont {Sandvik}},
  \bibinfo {author} {\bibfnamefont {A.}~\bibnamefont {Vertiatchikh}}, \bibinfo
  {author} {\bibfnamefont {J.}~\bibnamefont {Tucker}}, \ and\ \bibinfo {author}
  {\bibfnamefont {A.}~\bibnamefont {Elasser}},\ }\bibfield  {title} {\enquote
  {\bibinfo {title} {High temperature hall effect sensors based on
  {A}l{G}a{N}/{G}a{N} heterojunctions},}\ }\href@noop {} {\bibfield  {journal}
  {\bibinfo  {journal} {J. App. Phys.}\ }\textbf {\bibinfo {volume} {99}},\
  \bibinfo {pages} {114510--1--114510--4} (\bibinfo {year} {2006})}\BibitemShut
  {NoStop}%
\bibitem [{\citenamefont {Hout}\ and\ \citenamefont
  {Middelhoek}(1997)}]{hout97}%
  \BibitemOpen
  \bibfield  {author} {\bibinfo {author} {\bibfnamefont {S.~R.}\ \bibnamefont
  {Hout}}\ and\ \bibinfo {author} {\bibfnamefont {S.}~\bibnamefont
  {Middelhoek}},\ }\bibfield  {title} {\enquote {\bibinfo {title} {A 400\degree
  {C} silicon hall sensor},}\ }\href@noop {} {\bibfield  {journal} {\bibinfo
  {journal} {Sensors and Actuators A}\ }\textbf {\bibinfo {volume} {60}},\
  \bibinfo {pages} {14--22} (\bibinfo {year} {1997})}\BibitemShut {NoStop}%
\bibitem [{\citenamefont {Balint}\ \emph {et~al.}(2014)\citenamefont {Balint},
  \citenamefont {Cutts}, \citenamefont {Kolawa},\ and\ \citenamefont
  {Peterson}}]{caltech14}%
  \BibitemOpen
  \bibfield  {author} {\bibinfo {author} {\bibfnamefont {T.~S.}\ \bibnamefont
  {Balint}}, \bibinfo {author} {\bibfnamefont {J.~A.}\ \bibnamefont {Cutts}},
  \bibinfo {author} {\bibfnamefont {E.~A.}\ \bibnamefont {Kolawa}}, \ and\
  \bibinfo {author} {\bibfnamefont {C.~E.}\ \bibnamefont {Peterson}},\
  }\href@noop {} {\enquote {\bibinfo {title} {Extreme environment technologies
  for space and terrestrial applications},}\ }\bibinfo {type} {Tech. Rep.}\
  (\bibinfo  {institution} {JPL, California Institute of Technology},\ \bibinfo
  {year} {2014})\BibitemShut {NoStop}%
\bibitem [{\citenamefont {Maier}\ \emph {et~al.}(2012)\citenamefont {Maier},
  \citenamefont {Alomari}, \citenamefont {Grandjean}, \citenamefont {Carlin},
  \citenamefont {Diforte-Poisson}, \citenamefont {Dua}, \citenamefont
  {Delage},\ and\ \citenamefont {Kohn}}]{mai12}%
  \BibitemOpen
  \bibfield  {author} {\bibinfo {author} {\bibfnamefont {D.}~\bibnamefont
  {Maier}}, \bibinfo {author} {\bibfnamefont {M.}~\bibnamefont {Alomari}},
  \bibinfo {author} {\bibfnamefont {N.}~\bibnamefont {Grandjean}}, \bibinfo
  {author} {\bibfnamefont {J.-F.}\ \bibnamefont {Carlin}}, \bibinfo {author}
  {\bibfnamefont {M.-A.}\ \bibnamefont {Diforte-Poisson}}, \bibinfo {author}
  {\bibfnamefont {C.}~\bibnamefont {Dua}}, \bibinfo {author} {\bibfnamefont
  {S.}~\bibnamefont {Delage}}, \ and\ \bibinfo {author} {\bibfnamefont
  {E.}~\bibnamefont {Kohn}},\ }\bibfield  {title} {\enquote {\bibinfo {title}
  {In{A}l{N}/{G}a{N} {HEMT}s for operation in the 1000\degree {C} regime: A
  first experiment},}\ }\href@noop {} {\bibfield  {journal} {\bibinfo
  {journal} {IEEE Electron Dev. Lett.}\ }\textbf {\bibinfo {volume} {33}},\
  \bibinfo {pages} {985--987} (\bibinfo {year} {2012})}\BibitemShut {NoStop}%
\bibitem [{\citenamefont {Alpert}\ \emph {et~al.}(2019)\citenamefont {Alpert},
  \citenamefont {Dowling}, \citenamefont {Chapin}, \citenamefont {Yalamarthy},
  \citenamefont {Benbrook}, \citenamefont {K{\"{o}}ck}, \citenamefont
  {Ausserlechner},\ and\ \citenamefont {Senesky}}]{alp19}%
  \BibitemOpen
  \bibfield  {author} {\bibinfo {author} {\bibfnamefont {H.~S.}\ \bibnamefont
  {Alpert}}, \bibinfo {author} {\bibfnamefont {K.~M.}\ \bibnamefont {Dowling}},
  \bibinfo {author} {\bibfnamefont {C.~A.}\ \bibnamefont {Chapin}}, \bibinfo
  {author} {\bibfnamefont {A.~S.}\ \bibnamefont {Yalamarthy}}, \bibinfo
  {author} {\bibfnamefont {S.~R.}\ \bibnamefont {Benbrook}}, \bibinfo {author}
  {\bibfnamefont {H.}~\bibnamefont {K{\"{o}}ck}}, \bibinfo {author}
  {\bibfnamefont {U.}~\bibnamefont {Ausserlechner}}, \ and\ \bibinfo {author}
  {\bibfnamefont {D.~G.}\ \bibnamefont {Senesky}},\ }\bibfield  {title}
  {\enquote {\bibinfo {title} {Effect of geometry on sensitivity and offset of
  {A}l{G}a{N}/{G}a{N} and {I}n{A}l{N}/{G}a{N} hall-effect sensors},}\
  }\href@noop {} {\bibfield  {journal} {\bibinfo  {journal} {IEEE Sensors J.}\
  }\textbf {\bibinfo {volume} {19}},\ \bibinfo {pages} {3640--3646} (\bibinfo
  {year} {2019})}\BibitemShut {NoStop}%
\bibitem [{\citenamefont {Dowling}\ \emph {et~al.}(2019)\citenamefont
  {Dowling}, \citenamefont {Alpert}, \citenamefont {Yalamarthy}, \citenamefont
  {Satterthwaite}, \citenamefont {Kumar}, \citenamefont {K{\"{o}}ck},
  \citenamefont {Ausserlechner},\ and\ \citenamefont {Senesky}}]{dow19}%
  \BibitemOpen
  \bibfield  {author} {\bibinfo {author} {\bibfnamefont {K.~M.}\ \bibnamefont
  {Dowling}}, \bibinfo {author} {\bibfnamefont {H.~S.}\ \bibnamefont {Alpert}},
  \bibinfo {author} {\bibfnamefont {A.~S.}\ \bibnamefont {Yalamarthy}},
  \bibinfo {author} {\bibfnamefont {P.~F.}\ \bibnamefont {Satterthwaite}},
  \bibinfo {author} {\bibfnamefont {S.}~\bibnamefont {Kumar}}, \bibinfo
  {author} {\bibfnamefont {H.}~\bibnamefont {K{\"{o}}ck}}, \bibinfo {author}
  {\bibfnamefont {U.}~\bibnamefont {Ausserlechner}}, \ and\ \bibinfo {author}
  {\bibfnamefont {D.~G.}\ \bibnamefont {Senesky}},\ }\bibfield  {title}
  {\enquote {\bibinfo {title} {Micro-tesla offset in thermally stable
  {A}l{G}a{N}/{G}a{N} {2DEG} hall plates using current spinning},}\ }\href@noop
  {} {\bibfield  {journal} {\bibinfo  {journal} {IEEE Sensors Lett.}\ }\textbf
  {\bibinfo {volume} {3}} (\bibinfo {year} {2019})}\BibitemShut {NoStop}%
\bibitem [{\citenamefont {van~der Meer}\ \emph {et~al.}(2004)\citenamefont
  {van~der Meer}, \citenamefont {Riedijik}, \citenamefont {de~Jong},
  \citenamefont {van Kampen}, \citenamefont {Meekel},\ and\ \citenamefont
  {Huijsing}}]{vdm04}%
  \BibitemOpen
  \bibfield  {author} {\bibinfo {author} {\bibfnamefont {J.~C.}\ \bibnamefont
  {van~der Meer}}, \bibinfo {author} {\bibfnamefont {F.~R.}\ \bibnamefont
  {Riedijik}}, \bibinfo {author} {\bibfnamefont {P.~C.}\ \bibnamefont
  {de~Jong}}, \bibinfo {author} {\bibfnamefont {E.~A.}\ \bibnamefont {van
  Kampen}}, \bibinfo {author} {\bibfnamefont {M.~J.}\ \bibnamefont {Meekel}}, \
  and\ \bibinfo {author} {\bibfnamefont {J.~H.}\ \bibnamefont {Huijsing}},\
  }\bibfield  {title} {\enquote {\bibinfo {title} {{CMOS} quad spinning-current
  hall sensor system for compass application},}\ }in\ \href@noop {} {\emph
  {\bibinfo {booktitle} {Proc. Int. Conf. IEEE Sensors}}}\ (\bibinfo {year}
  {2004})\ pp.\ \bibinfo {pages} {1434--1437}\BibitemShut {NoStop}%
\bibitem [{\citenamefont {van~der Meer}\ \emph {et~al.}(2005)\citenamefont
  {van~der Meer}, \citenamefont {Riedijik}, \citenamefont {van Kampen},
  \citenamefont {Makinwa},\ and\ \citenamefont {Huijsing}}]{vdm05}%
  \BibitemOpen
  \bibfield  {author} {\bibinfo {author} {\bibfnamefont {J.~C.}\ \bibnamefont
  {van~der Meer}}, \bibinfo {author} {\bibfnamefont {F.~R.}\ \bibnamefont
  {Riedijik}}, \bibinfo {author} {\bibfnamefont {E.}~\bibnamefont {van
  Kampen}}, \bibinfo {author} {\bibfnamefont {K.~A.~A.}\ \bibnamefont
  {Makinwa}}, \ and\ \bibinfo {author} {\bibfnamefont {J.~H.}\ \bibnamefont
  {Huijsing}},\ }\bibfield  {title} {\enquote {\bibinfo {title} {A fully
  integrated {CMOS} hall sensor with a 3.65 $\mu${T} 3 $\sigma$ offset for
  compass applications},}\ }in\ \href@noop {} {\emph {\bibinfo {booktitle}
  {Proc. Int. Conf. IEEE Sensors}}}\ (\bibinfo {address} {San Francisco, CA,
  USA},\ \bibinfo {year} {2005})\ pp.\ \bibinfo {pages} {246--247}\BibitemShut
  {NoStop}%
\bibitem [{\citenamefont {Ausserlechner}(2016{\natexlab{a}})}]{aus16}%
  \BibitemOpen
  \bibfield  {author} {\bibinfo {author} {\bibfnamefont {U.}~\bibnamefont
  {Ausserlechner}},\ }\bibfield  {title} {\enquote {\bibinfo {title} {Hall
  effect devices with three terminals: Their magnetic sensitivity and offset
  cancellation scheme},}\ }\href@noop {} {\bibfield  {journal} {\bibinfo
  {journal} {J. of Sensors}\ }\textbf {\bibinfo {volume} {2016}} (\bibinfo
  {year} {2016}{\natexlab{a}})}\BibitemShut {NoStop}%
\bibitem [{\citenamefont {Ruther}\ \emph {et~al.}(2003)\citenamefont {Ruther},
  \citenamefont {Schiller}, \citenamefont {Janke},\ and\ \citenamefont
  {Paul}}]{ruth03}%
  \BibitemOpen
  \bibfield  {author} {\bibinfo {author} {\bibfnamefont {P.}~\bibnamefont
  {Ruther}}, \bibinfo {author} {\bibfnamefont {U.}~\bibnamefont {Schiller}},
  \bibinfo {author} {\bibfnamefont {R.}~\bibnamefont {Janke}}, \ and\ \bibinfo
  {author} {\bibfnamefont {O.}~\bibnamefont {Paul}},\ }\bibfield  {title}
  {\enquote {\bibinfo {title} {Thermomagnetic residual offset in integrated
  hall plates},}\ }\href@noop {} {\bibfield  {journal} {\bibinfo  {journal}
  {IEEE Sensors J.}\ }\textbf {\bibinfo {volume} {3}},\ \bibinfo {pages}
  {693--699} (\bibinfo {year} {2003})}\BibitemShut {NoStop}%
\bibitem [{\citenamefont {Randhawa}(1981)}]{rand81}%
  \BibitemOpen
  \bibfield  {author} {\bibinfo {author} {\bibfnamefont {G.~S.}\ \bibnamefont
  {Randhawa}},\ }\bibfield  {title} {\enquote {\bibinfo {title} {Monolithic
  integrated hall devices in silicon circuits},}\ }\href@noop {} {\bibfield
  {journal} {\bibinfo  {journal} {Microelectronics J.}\ }\textbf {\bibinfo
  {volume} {12}},\ \bibinfo {pages} {24--29} (\bibinfo {year}
  {1981})}\BibitemShut {NoStop}%
\bibitem [{\citenamefont {Sander}\ \emph {et~al.}(2015)\citenamefont {Sander},
  \citenamefont {Leube}, \citenamefont {Aftab}, \citenamefont {Ruther},\ and\
  \citenamefont {Paul}}]{san15}%
  \BibitemOpen
  \bibfield  {author} {\bibinfo {author} {\bibfnamefont {C.}~\bibnamefont
  {Sander}}, \bibinfo {author} {\bibfnamefont {C.}~\bibnamefont {Leube}},
  \bibinfo {author} {\bibfnamefont {T.}~\bibnamefont {Aftab}}, \bibinfo
  {author} {\bibfnamefont {P.}~\bibnamefont {Ruther}}, \ and\ \bibinfo {author}
  {\bibfnamefont {O.}~\bibnamefont {Paul}},\ }\bibfield  {title} {\enquote
  {\bibinfo {title} {Isotropic 3{D} silicon hall sensor},}\ }in\ \href@noop {}
  {\emph {\bibinfo {booktitle} {Proc. 28th Int. Conf. IEEE Micro. Electro.
  Mech. Syst.}}}\ (\bibinfo {address} {Estoril, Portugal},\ \bibinfo {year}
  {2015})\ pp.\ \bibinfo {pages} {893--896}\BibitemShut {NoStop}%
\bibitem [{\citenamefont {Bouguen}\ \emph {et~al.}(2009)\citenamefont
  {Bouguen}, \citenamefont {Konczewicz}, \citenamefont {Contreras},
  \citenamefont {Jouault}, \citenamefont {Camassel},\ and\ \citenamefont
  {Cordier}}]{boug09}%
  \BibitemOpen
  \bibfield  {author} {\bibinfo {author} {\bibfnamefont {L.}~\bibnamefont
  {Bouguen}}, \bibinfo {author} {\bibfnamefont {L.}~\bibnamefont {Konczewicz}},
  \bibinfo {author} {\bibfnamefont {S.}~\bibnamefont {Contreras}}, \bibinfo
  {author} {\bibfnamefont {B.}~\bibnamefont {Jouault}}, \bibinfo {author}
  {\bibfnamefont {J.}~\bibnamefont {Camassel}}, \ and\ \bibinfo {author}
  {\bibfnamefont {Y.}~\bibnamefont {Cordier}},\ }\bibfield  {title} {\enquote
  {\bibinfo {title} {High temperature behaviour of {A}l{G}a{N}/{G}a{N}
  hall-{FET} sensors},}\ }\href@noop {} {\bibfield  {journal} {\bibinfo
  {journal} {Materials Sci. and Eng. B}\ }\textbf {\bibinfo {volume} {165}},\
  \bibinfo {pages} {1--4} (\bibinfo {year} {2009})}\BibitemShut {NoStop}%
\bibitem [{\citenamefont {Abderrahmane}\ \emph {et~al.}(2012)\citenamefont
  {Abderrahmane}, \citenamefont {Koide}, \citenamefont {Sato}, \citenamefont
  {Ohshima}, \citenamefont {Sandhu},\ and\ \citenamefont {Okada}}]{abd12}%
  \BibitemOpen
  \bibfield  {author} {\bibinfo {author} {\bibfnamefont {A.}~\bibnamefont
  {Abderrahmane}}, \bibinfo {author} {\bibfnamefont {S.}~\bibnamefont {Koide}},
  \bibinfo {author} {\bibfnamefont {S.-I.}\ \bibnamefont {Sato}}, \bibinfo
  {author} {\bibfnamefont {T.}~\bibnamefont {Ohshima}}, \bibinfo {author}
  {\bibfnamefont {A.}~\bibnamefont {Sandhu}}, \ and\ \bibinfo {author}
  {\bibfnamefont {H.}~\bibnamefont {Okada}},\ }\bibfield  {title} {\enquote
  {\bibinfo {title} {Robust hall effect magnetic field sensors for operation at
  high temperatures and in harsh radiation environments},}\ }\href@noop {}
  {\bibfield  {journal} {\bibinfo  {journal} {IEEE Trans. on Magnetics}\
  }\textbf {\bibinfo {volume} {48}},\ \bibinfo {pages} {4421--4423} (\bibinfo
  {year} {2012})}\BibitemShut {NoStop}%
\bibitem [{\citenamefont {Wang}\ \emph {et~al.}(2007)\citenamefont {Wang},
  \citenamefont {Shen}, \citenamefont {Xu}, \citenamefont {Wang}, \citenamefont
  {Xu}, \citenamefont {Huang}, \citenamefont {Yang}, \citenamefont {Xu},\ and\
  \citenamefont {Zhang}}]{wang07}%
  \BibitemOpen
  \bibfield  {author} {\bibinfo {author} {\bibfnamefont {M.~J.}\ \bibnamefont
  {Wang}}, \bibinfo {author} {\bibfnamefont {B.}~\bibnamefont {Shen}}, \bibinfo
  {author} {\bibfnamefont {F.~J.}\ \bibnamefont {Xu}}, \bibinfo {author}
  {\bibfnamefont {Y.}~\bibnamefont {Wang}}, \bibinfo {author} {\bibfnamefont
  {J.}~\bibnamefont {Xu}}, \bibinfo {author} {\bibfnamefont {S.}~\bibnamefont
  {Huang}}, \bibinfo {author} {\bibfnamefont {Z.~J.}\ \bibnamefont {Yang}},
  \bibinfo {author} {\bibfnamefont {K.}~\bibnamefont {Xu}}, \ and\ \bibinfo
  {author} {\bibfnamefont {G.~Y.}\ \bibnamefont {Zhang}},\ }\bibfield  {title}
  {\enquote {\bibinfo {title} {High temperature dependence of the density of
  two-dimensional electron gas in {A}l$_{0.18}${G}a$_{0.82}${N}/{G}a{N}
  heterostructures},}\ }\href@noop {} {\bibfield  {journal} {\bibinfo
  {journal} {Appl. Phys. A}\ ,\ \bibinfo {pages} {715--718}} (\bibinfo {year}
  {2007})}\BibitemShut {NoStop}%
\bibitem [{\citenamefont {Ridley}, \citenamefont {Foutz},\ and\ \citenamefont
  {Eastman}(2000)}]{rid00}%
  \BibitemOpen
  \bibfield  {author} {\bibinfo {author} {\bibfnamefont {B.~K.}\ \bibnamefont
  {Ridley}}, \bibinfo {author} {\bibfnamefont {B.~E.}\ \bibnamefont {Foutz}}, \
  and\ \bibinfo {author} {\bibfnamefont {L.~F.}\ \bibnamefont {Eastman}},\
  }\bibfield  {title} {\enquote {\bibinfo {title} {Mobility of electrons in
  bulk {G}a{N} and {A}l$_x${G}a$_{1-x}${N}/{G}a{N} heterostructures},}\
  }\href@noop {} {\bibfield  {journal} {\bibinfo  {journal} {Phys. Rev. B,
  Condens. Matter}\ }\textbf {\bibinfo {volume} {611}},\ \bibinfo {pages}
  {16862--16869} (\bibinfo {year} {2000})}\BibitemShut {NoStop}%
\bibitem [{\citenamefont {Ausserlechner}(2017)}]{aus17}%
  \BibitemOpen
  \bibfield  {author} {\bibinfo {author} {\bibfnamefont {U.}~\bibnamefont
  {Ausserlechner}},\ }\bibfield  {title} {\enquote {\bibinfo {title} {The
  signal-to-noise ratio and a hidden symmetry of hall plates},}\ }\href@noop {}
  {\bibfield  {journal} {\bibinfo  {journal} {Solid-State Electron.}\ }\textbf
  {\bibinfo {volume} {135}},\ \bibinfo {pages} {14--23} (\bibinfo {year}
  {2017})}\BibitemShut {NoStop}%
\bibitem [{\citenamefont {Popovi\'{c}}(2004)}]{popo04}%
  \BibitemOpen
  \bibfield  {author} {\bibinfo {author} {\bibfnamefont {R.}~\bibnamefont
  {Popovi\'{c}}},\ }\href@noop {} {\emph {\bibinfo {title} {Hall Effect
  Devices}}},\ \bibinfo {edition} {2nd}\ ed.\ (\bibinfo  {publisher} {IOP},\
  \bibinfo {address} {UK},\ \bibinfo {year} {2004})\BibitemShut {NoStop}%
\bibitem [{\citenamefont {Ausserlechner}(2016{\natexlab{b}})}]{aus16_2}%
  \BibitemOpen
  \bibfield  {author} {\bibinfo {author} {\bibfnamefont {U.}~\bibnamefont
  {Ausserlechner}},\ }\bibfield  {title} {\enquote {\bibinfo {title} {Closed
  form expressions for sheet resistance and mobility from van-der-pauw
  measurement on 90\degree symmetric devices with four arbitrary contacts},}\
  }\href@noop {} {\bibfield  {journal} {\bibinfo  {journal} {Solid-State
  Electron.}\ }\textbf {\bibinfo {volume} {116}},\ \bibinfo {pages} {46--54}
  (\bibinfo {year} {2016}{\natexlab{b}})}\BibitemShut {NoStop}%
\bibitem [{\citenamefont {Aminbeidokhti}\ \emph {et~al.}(2016)\citenamefont
  {Aminbeidokhti}, \citenamefont {Dimitrijev}, \citenamefont {Han},
  \citenamefont {Chen},\ and\ \citenamefont {Xu}}]{amin16}%
  \BibitemOpen
  \bibfield  {author} {\bibinfo {author} {\bibfnamefont {A.}~\bibnamefont
  {Aminbeidokhti}}, \bibinfo {author} {\bibfnamefont {S.}~\bibnamefont
  {Dimitrijev}}, \bibinfo {author} {\bibfnamefont {J.}~\bibnamefont {Han}},
  \bibinfo {author} {\bibfnamefont {X.}~\bibnamefont {Chen}}, \ and\ \bibinfo
  {author} {\bibfnamefont {X.}~\bibnamefont {Xu}},\ }\bibfield  {title}
  {\enquote {\bibinfo {title} {The power law of phonon-limited electron
  mobility in the 2-{D} electron gas of {A}l{G}a{N}/{G}a{N} heterostructure},}\
  }\href@noop {} {\bibfield  {journal} {\bibinfo  {journal} {IEEE Trans.
  Electron. Dev.}\ }\textbf {\bibinfo {volume} {63}},\ \bibinfo {pages}
  {2214--2218} (\bibinfo {year} {2016})}\BibitemShut {NoStop}%
\bibitem [{\citenamefont {Mnatsakanov}\ \emph {et~al.}(2003)\citenamefont
  {Mnatsakanov}, \citenamefont {Levinshtein}, \citenamefont {Pomortseva},
  \citenamefont {Yurkov}, \citenamefont {Simin},\ and\ \citenamefont
  {Khan}}]{mnat03}%
  \BibitemOpen
  \bibfield  {author} {\bibinfo {author} {\bibfnamefont {T.~T.}\ \bibnamefont
  {Mnatsakanov}}, \bibinfo {author} {\bibfnamefont {M.~E.}\ \bibnamefont
  {Levinshtein}}, \bibinfo {author} {\bibfnamefont {L.~I.}\ \bibnamefont
  {Pomortseva}}, \bibinfo {author} {\bibfnamefont {S.~N.}\ \bibnamefont
  {Yurkov}}, \bibinfo {author} {\bibfnamefont {G.~S.}\ \bibnamefont {Simin}}, \
  and\ \bibinfo {author} {\bibfnamefont {M.~A.}\ \bibnamefont {Khan}},\
  }\bibfield  {title} {\enquote {\bibinfo {title} {Carrier mobility model for
  {G}a{N}},}\ }\href@noop {} {\bibfield  {journal} {\bibinfo  {journal}
  {Solid-State Electron.}\ }\textbf {\bibinfo {volume} {47}},\ \bibinfo {pages}
  {111--115} (\bibinfo {year} {2003})}\BibitemShut {NoStop}%
\bibitem [{\citenamefont {Ausserlechner}, \citenamefont {Motz},\ and\
  \citenamefont {Holliber}(2004)}]{aus04}%
  \BibitemOpen
  \bibfield  {author} {\bibinfo {author} {\bibfnamefont {U.}~\bibnamefont
  {Ausserlechner}}, \bibinfo {author} {\bibfnamefont {M.}~\bibnamefont {Motz}},
  \ and\ \bibinfo {author} {\bibfnamefont {M.}~\bibnamefont {Holliber}},\
  }\bibfield  {title} {\enquote {\bibinfo {title} {Drift of magnetic
  sensitivity of smart {H}all sensors due to moisture absorbed by the
  {IC}-package [automotive applications]},}\ }in\ \href@noop {} {\emph
  {\bibinfo {booktitle} {SENSORS}}}\ (\bibinfo {address} {Vienna, Austria},\
  \bibinfo {year} {2004})\ pp.\ \bibinfo {pages} {455--458}\BibitemShut
  {NoStop}%
\end{thebibliography}%

\end{document}